\documentclass[10pt]{article}
\usepackage{a4wide}
\usepackage{epsfig}
\usepackage{amsmath}
\usepackage[T1]{fontenc}
\renewcommand{\Re}{\operatorname{Re}}
\newcommand{\li}{\operatorname{Li_2}}
\newtheorem{lemma}{Lemma}

\author{Richard Kreckel\footnote{e-mail: {\tt richard.kreckel@uni-mainz.de}}, 
Dirk Kreimer\footnote{e-mail: {\tt dirk.kreimer@uni-mainz.de}, Heisenberg Fellow of the DFG}, 
Karl Schilcher\footnote{e-mail: {\tt karl@thep.physik.uni-mainz.de}}}

\date{April 20, 1998}
\title{\begin{flushright}
\null
\vskip-3cm
{\normalsize MZ-TH/98-16}
\vskip 1cm
\end{flushright}
First Results with a new Method for calculating 2-loop Box-Functions}

\begin{document}
\maketitle

\begin{abstract}
  We describe a first attempt to calculate scalar 2-loop box-functions
  with arbitrary internal masses, applying a novel method proposed
  in~\cite{hepph9407234}. Four of the eight integrals are accessible to
  integration by means of the residue theorem, leaving a rational
  function in the remaining variables. The result of the procedure is
  a three- or sometimes two-dimensional integral representation over a
  finite volume that can be further evaluated using numerical methods.
\end{abstract}

The study of higher-loop quantum corrections in QFTs is accompanied by
rapidly increasing computational challenges. While in pure QED e.g.\
the presence of only one mass scale allows relative high loop orders
to be evaluated~\cite{KinoshitaQED}, the feasibility of
loop calculations in the general case with arbitrary internal masses
has up to now been cut off at the two-loop level.

\begin{figure}
 \begin{center} \epsfig{figure=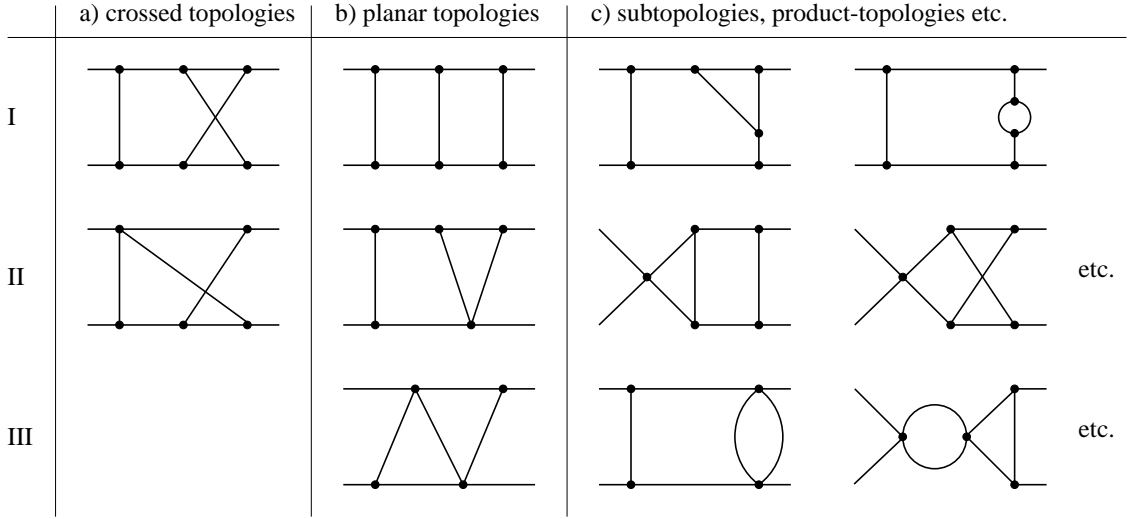,width=\textwidth}
 \end{center} 
\caption{Possible 2-loop box-topologies. Starting from class~I, where all
topologies are constructed by appending 4 external legs to the 2-loop 
master-topology, one can construct all other box-topologies by successive
cancellation of internal propagators.}
 \label{fig:functions}
\end{figure}

In recent years much work has been done on two-loop integrals with
arbitrary internal masses of the self-energy- and vertex-type. These
functions could be reduced to double integrals with manageable
numerical behaviour~\cite{MasterTwo-Two, MasterTwo-Three, NewRep,
  NewRepCrossed} (c.f.~\cite{DavTausk, FleischerTentyukov} for another
approach at these diagrams). Not much seems to be known, however,
about the different two-loop corrections to scattering-amplitudes
(Fig.~\ref{fig:functions}) involving arbitrary-mass propagators which
will become of some importance for probing new physics in
\(W^+W^-\)-processes or in a Muon-collider, for example.  Some of
these topologies factorize into products of one-loop functions, i.e.\ 
no propagator shares both loop-momenta.  Others are one-loop functions
containing other one-loop functions as subtopologies
(Fig.~\ref{fig:functions}, third column). Rather than analyzing all of
these functions, we focus our attention on five four-point topologies.
They are all obtained by shrinking propagators in the two basic
four-point topologies \epsfig{figure=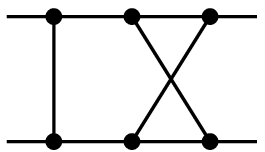,height=2ex} and
\epsfig{figure=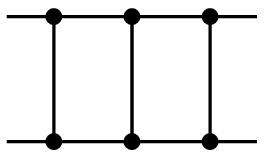,height=2ex}. Two of them are of
crossed type where two propagators share both loop-momenta (first
column) and three of planar type (second column).  There is no
striking difference to the graphs in the third column. For example the
first graph in the first row, third column
\epsfig{figure=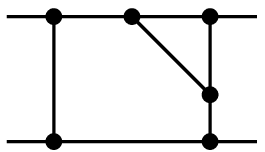,height=2ex} relates to the
graph in the second row, second column
\epsfig{figure=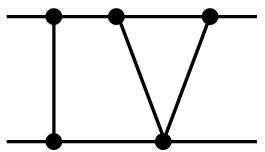,height=2ex} if we shrink the appropriate
propagator.

The importance of the four point function derives not only from
the fact that it describes the first scattering amplitude
(2 \(\rightarrow\) 2), but also from the fact that the multiparticle
scattering amplitudes, involving \(n\)-point functions, \(n\geq 5\),
are closely related to the four point functions \cite{nickel}.

We will see below how this relation emerges in our method at the
two-loop level.

The idea behind our method is to perform four of the integrals with
Cauchy's residue theorem and evaluate at least one of the remaining
ones analytically, leaving a two- or three-dimensional representation
for numerical evaluation.

This article is organized as follows: In the first two sections we
describe how one can generally perform four integrations with the
residue theorem alone and analyze the structure of the result. In the
remaining sections we will complete the calculation for a graph of
type \epsfig{figure=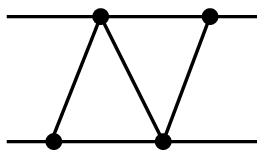,height=2ex} in a special kinematic
case, always keeping in mind the more general graphs, notably
\epsfig{figure=ladder.eps,height=2ex} and
\epsfig{figure=crossedladder.eps,height=2ex} as we go along. In
section~\ref{sec:example} we look at some of the results and discuss
what possible problems we need to be concerned with in some
limiting kinematical regimes.

\section{Integrating the middle variables}

The functions under consideration can generally be written as
\begin{equation}
V = \int d^4k\int d^4l\,\frac{1}{P_{l,m_1}\,P_{l+k,m_2}\,P_{k,m_3}\cdots},
\end{equation}
with five to seven inverse scalar propagators of the form
\(P_{l,m_i}=(l+p)^2-m_i^2+i\eta\), \(l\) denoting a loop-momentum
and \(p\) some combination of external 4-momenta. Note that we do not
attach an index \(i\) to the \(\eta\) from the Feynman prescription
since we will choose them all to be equal.  \(V\) depends on six
independent kinematic variables, a possible choice being the
Mandelstam-variables \(s\), \(t\) and \(u\) together with the masses
\(m_i\) of the four external particles and the condition
\(s+t+u=\sum_i m_i^2\). We will, however, need to choose an explicit
Lorentz frame for our purposes.

When trying to perform one of the integrals using the residue theorem,
it is very attractive to first linearize all the propagators \(P_i\)
with respect to the corresponding variable since this makes the
detection of poles particularly simple and does not introduce
unnecessary square roots inhibiting further integrations. Due to the
signature of the Minkowski metric this linearization can easily be
done in one pair of variables. Choosing \(l_1\) and \(k_1\) for
linearization and applying the shift
\begin{equation} \label{shiftone}
l_0 \longrightarrow l_0 + l_1,\qquad k_0 \longrightarrow k_0 + k_1,
\end{equation}
a propagator \(P_{l}\) undergoes the transformation
\begin{eqnarray} \label{shifteffect}
(l+p)^2 - m^2 + i\eta = (l_0+p_0)^2 - (l_1+p_1)^2 -(l_2+p_2)^2 - (l_3+p_3)^2 - m^2 +i\eta\, \nonumber\\
 \longrightarrow (l_0+p_0)^2 + 2\,l_1\,(l_0+p_0-p_1) - p_1^2 - (l_2+p_2)^2 - (l_3+p_3)^2 - m^2 +i\eta.
\end{eqnarray}
This shift can safely be done since the functions defined by topologies 
I\,a), II\,a), and I\,b) - III\,b) converge absolutely as can be seen by 
counting powers.

Closing the contour in the upper half-plane and using the residue
theorem to carry out the \(l_1\)- and \(k_1\)-integrations, one
obtains constraints for the \(l_0\)- and \(k_0\)-integrations.  This
appearance of constraints is due to the position of poles in the
complex \(l_1\)- and \(k_1\)-planes either in the upper or the lower
half and hence either contributing as a residue or not. These
constraints affect the \(l_0\)- and \(k_0\)-integrations only because
no other loop-variables appear in the \(l_1\)-linear term
in~(\ref{shifteffect}). We will return to these constraints in the
next section.

Noticing that the integrations over the variables \(l_2\), \(l_3\),
\(k_2\) and \(k_3\) are still unbounded, suggests solving two of them
with the residue theorem again. In~\cite{hepph9407234} it was shown
how the linearization necessary for simplifying this task can be
carried out in what we may call the middle variables: \(k_1\),
\(l_1\), \(k_2\) and \(l_2\). We will briefly repeat the argument.

As a consequence of conservation of four-momentum, the external legs
of any four-point function span a 3-dimensional subspace of
momentum-space---the so-called parallel space. Using
Lorentz-invariance, its complement---the orthogonal space---can always
be chosen to be parallel to the 3-axis. The 3-components of the
loop-momenta do not mix with any external momenta then:

\begin{eqnarray}
P_{l} &=&l_3^2 + (\textit{something real}) + i\eta, \nonumber\\
P_{l+k}&=&(l_3+k_3)^2 + (\textit{something real}) + i\eta, \nonumber\\
P_{k}&=&k_3^2 + (\textit{something real}) + i\eta. \nonumber
\end{eqnarray}
Hence the poles of the integrand in the complex \(l_3\), \(k_3\) and
\((l_3+k_3)\)-planes respectively are all located in the first and third
quadrant. 

At this stage one can make contact to the case of a general
\(n\)-point function. For \(n\geq 5\) the 3-components do mix with
external momenta. But the very fact that \(n\geq 5\) ensures that {\em
  after} undertaking appropriate partial fraction in the propagators
the convergence of all integrals is sufficiently strong so that
termwise linear shifts are allowed which erase the appearance of
external momenta in the 3-components of all quadratic propagators.

The observation that the poles of the integrand are in the first and
third quadrant (together with the fact that the integrand falls off
sufficiently rapidly at large \(l_3\), \(k_3\) and \(l_3+k_3\))
suggests that we should try to rotate clockwise by \(\pi/2\) and
effectively change the metrics from the usual Minkowski metric
\((+,-,-,-)\) to \((+,-,-,+)\) in order to be able to linearize the
integrand in \(l_2\) and \(k_2\). The mixed propagator \(P_{l+k}\),
however, seems to spoil this project since its roots in the complex
\(k_3\)- or \(l_3\)-plane alone are not bound to the first and third
quadrant. In order to do the rotation, we have to treat \(l_3\),
\(k_3\) and \(l_3 + k_3\) on the same footing. Due to the integrand's
symmetry we can restrict our attention to the first quadrant in the
\(k_3\)-\(l_3\)-plane:
\begin{eqnarray}
  \int_{-\infty}^{+\infty}\!\!dl_3\int_{-\infty}^{+\infty}\!\!dk_3\,
  \frac{1}{P_{l}\bigl(l_3^2\bigr)\,P_{l+k}\bigl((l_3+k_3)^2\bigr)\,P_{k}\bigl(k_3^2\bigr)\cdots}\qquad \nonumber\\
  = 2 \int_0^{+\infty}\!\!dl_3\int_0^{+\infty}\!\!dk_3
  \left(\frac{1}{P_{l}\bigl(l_3^2\bigr)\,P_{l+k}\bigl((l_3+k_3)^2\bigr)\,P_{k}\bigl(k_3^2\bigr)\cdots}\,\,\,\right. \nonumber\\
  \left.{}+\frac{1}{P_{l}(l_3^2)\,P_{l+k}\bigl((l_3-k_3)^2\bigr)\,P_{k}\bigl(k_3^2\bigr)\cdots}\right)
  \nonumber
\end{eqnarray}
Now we reparametrize this quadrant by substituting \(l_3^2
\rightarrow u\,v^2\) and \(k_3^2 \rightarrow (1-u)\,v^2\):
\begin{eqnarray}
  \frac{1}{2}\int_{0}^{1}\!\!\frac{du}{\sqrt{u\,(1-u)}}\int_{0}^{+\infty}\!\!v\,dv
  \left(\frac{1}{P_{l}\bigl(u\,v^2\bigr)\,P_{l+k}\bigl((\sqrt{u}+\sqrt{1-u}\bigr)^2\,v^2)\,P_{k}\bigl((1-u)\,v^2\bigr)\cdots}\,\,\,\right. \nonumber\\
  \left.{}+\frac{1}{P_{l}\bigl(u\,v^2\bigr)\,P_{l+k}\bigl((\sqrt{u}-\sqrt{1-u})^2\,v^2\bigr)\,P_{k}\bigl((1-u)\,v^2\bigr)\cdots}\right) \nonumber
\end{eqnarray}
Here, \(u\), \((1-u)\) and \((\sqrt{u}-\sqrt{1-u})^2\) are all
positive so the propagators have their poles in the first and third
quadrant of the complex \(v\)-plane. This allows us to close the
contour of \(v\)-integration around the fourth quadrant, with the
Jacobian picking up a sign:
\begin{eqnarray}
  \frac{1}{2}\int_{0}^{1}\!\!\frac{du}{\sqrt{u\,(1-u)}}\int_{0}^{+\infty}\!\!\!\!-v\,dv
  \left(\frac{1}{P_{l}\bigl(-u\,v^2\bigr)\,P_{l+k}\bigl(-(\sqrt{u}+\sqrt{1-u})^2\,v^2\bigr)\,P_{k}\bigl(-(1-u)\,v^2\bigr)\cdots}\quad\right. \nonumber\\
  \left.{}+\frac{1}{P_{l}\bigl(-u\,v^2\bigr)\,P_{l+k}\bigl(-(\sqrt{u}-\sqrt{1-u})^2\,v^2\bigr)\,P_{k}\bigl(-(1-u)\,v^2\bigr)\cdots}\right). \nonumber
\end{eqnarray}
Inverting the above transformations we are left with the identity
\begin{eqnarray} \label{KreimerTrafo}
 \lefteqn{\int_{-\infty}^{+\infty}\!\!dl_3\int_{-\infty}^{+\infty}\!\!dk_3\,
  \frac{1}{P_{l}\bigl(l_3^2\bigr)\,P_{l+k}\bigl((l_3+k_3)^2\bigr)\,P_{k}\bigl(k_3^2\bigr)\cdots}} \nonumber\\
  & = & -\int_{-\infty}^{+\infty}\!\!dl_3\int_{-\infty}^{+\infty}\!\!dk_3\,
  \frac{1}{P_{l}\bigl(-l_3^2\bigr)\,P_{l+k}\bigl(-(l_3+k_3)^2\bigr)\,P_{k}\bigl(-k_3^2\bigr)\cdots}.
\end{eqnarray}
Note that this flip in the sign of metric is not related to the
standard Wick rotation~\cite{ItzZub} where an appropriate analytic
continuation has to be applied at the end of calculation in order to
obtain the Greens function for arbitrary exterior momenta. The
flip~(\ref{KreimerTrafo}) is a simple analytical formula not for our
case only but for any loop-component belonging to orthogonal space. In
particular, we are not allowed to set the imaginary part in the
propagators to zero.

Now we are in a position to complete the linearization of our
propagators in the middle variables \(k_1\), \(l_1\), \(k_2\) and
\(l_2\) by applying the shifts
\begin{equation} \label{shifttwo}
l_3 \longrightarrow l_3 + l_2,\qquad k_3 \longrightarrow k_3 + k_2
\end{equation}
in addition to~(\ref{shiftone}).

Next we interchange the order of integration in order to apply the
residue theorem to the four middle variables first. Recall that we are
allowed to interchange the order of integration for our graphs because
the integral over the modulus of their integrand exist. 

As mentioned above, the sign of the linear coefficient of the variable
being integrated (\(l_1\) in~(\ref{shifteffect})) determines whether the
pole is inside or outside the contour. We therefore obtain a sum of
residues each having a Heaviside function constraining the domain of
integration of the edge variables \(k_0\), \(l_0\), \(k_3\) and
\(l_3\) arising at each integration of the middle variables.

\section{Constraints}
\label{sec:constraints}

At each integration, we expect to obtain one residue for each
propagator containing the integration-variable. The potential
proliferation of terms is, however, drastically reduced by two
different kinds of relations holding among them as we show next.

The first relation is a well-known corollary to the residue theorem
(see e.g.~\cite{CartanAnalyticFunctions}):
\begin{lemma}
  The sum of the residues of a rational function (including a possible
  residue at infinity) is zero.
\end{lemma}
Therefore we can express one of the remaining residues after each
integration by the sum of all the other ones. In our case we apply
this to a product of inverse propagators linear in an
integration-variable \(l\): \(P_{l,i}=\alpha_i+\beta_i l+i\eta\), with
\(\eta > 0\), \(n>1\) and \(\alpha_i\) containing the remaining mass-
and momentum-terms.  We are allowed to drop one term and only keep its
\(\theta\)-function:
\begin{eqnarray}
\frac{1}{2 \pi i}\int_{-\infty}^{+\infty}\!\!dl\,\frac{1}{P_{l,1}\,P_{l,2}} & = &
\frac{\theta(-\beta_1)-\theta(-\beta_2)}{P_{l,2}|_{l^{(1)}}\,\beta_1}
\label{boxkorollar2} \nonumber\\
\frac{1}{2 \pi i}\int_{-\infty}^{+\infty}\!\!dl\,\frac{1}{P_{l,1}\,P_{l,2}\,P_{l,3}} & = &
\frac{\theta(-\beta_1)-\theta(-\beta_3)}{P_{l,2}|_{l^{(1)}}\,P_{l,3}|_{l^{(1)}}\,\beta_1}+ 
\frac{\theta(-\beta_2)-\theta(-\beta_3)}{P_{l,1}|_{l^{(2)}}\,P_{l,3}|_{l^{(2)}}\,\beta_2} 
\label{boxkorollar3} \nonumber\\
& \vdots & \nonumber \\
\frac{1}{2 \pi i}\int_{-\infty}^{+\infty}\!\!dl \prod_{i=1}^n \frac{1}{P_{l,i}} & = &
\sum_{i=1}^{n-1}\frac{\theta(-\beta_i)-\theta(-\beta_n)}{\prod_{j \neq i}P_{l,j}|_{l^{(i)}}\,\beta_i}
\end{eqnarray}
where \(l^{(i)}\) denotes the zero of \(P_{l,i}\). Since this
collecting can be done in four consecutive integrations, one can reduce
the number of terms for the planar box-function
\epsfig{figure=ladder.eps,height=2ex} from 108 to 36.

The second relation holding among the terms is a consequence of the
consecutive integration in \(l_1\) and \(l_2\) and can be stated as
follows:
\begin{lemma}
  Consider the term obtained by evaluating first the residue at the
  pole \(l_1^{(i)}\) due to \(P_{l,i}\) in the \(l_1\)-integration and
  then the residue at the pole \(l_2^{(j)}\) due to \(P_{l,j}\) (with
  \(l_1^{(i)}\) inserted) in the \(l_2\)-integration. It differs from
  the one obtained by first calculating the residue due to \(P_{l,j}\)
  and then to \(P_{l,i}\) (with \(l_1^{(j)}\) inserted) by a sign
  only.
\end{lemma}
To prove it, we write the inverse propagators as
\(P_{l,i}=\alpha_i+\beta_{i1}l_1+\beta_{i2}l_2\) and note that the
locations of poles \(l_1^{(i)}\), \(l_1^{(j)}\), \(l_2^{(i)}\) and
\(l_2^{(j)}\) are obtained by solving a linear system
\begin{equation} \label{LemmaTwoSystem}
 \left(\begin{array}{c} P_{l,i} \\ P_{l,j} \end{array} \right) =
 \left(\begin{array}{c} \alpha_i \\ \alpha_j \end{array} \right) +
 \left(\begin{array}{cc} \beta_{i1} & \beta_{i2} \\ 
                         \beta_{j1} & \beta_{j2} \\ \end{array} \right) 
 \left(\begin{array}{c} l_1 \\ l_2 \end{array} \right) \stackrel{!}{=} 
 \left(\begin{array}{c} 0 \\ 0 \end{array} \right).
\end{equation}
Hence the two orders of integration amount to the two ways of solving
such a system: first solving the first line and inserting it into the
second or vice versa. The solutions inserted into the remaining
\(P_q\) are therefore the same: \(l_1^{(i)}=l_1^{(j)}\) and
\(l_2^{(i)}=l_2^{(j)}\). This, however, specifies only proportionality
in our second lemma. The proportionality-constant \(-1\) can readily
be obtained by writing down the residual term after evaluating the
\((1,i)\rightarrow (2,j)\)-order:
\begin{displaymath}
V_{[i,j]}=\frac{1}{(\beta_{i1}\beta_{j2} - \beta_{j1}\beta_{i2})\,P_{l,q}|_{l_1^{(i)},l_2^{(j)}}\cdots},
\end{displaymath}
which is antisymmetric in \(i\) and \(j\).

In the planar case, Lemma 2 can always be applied twice to
restrict the number of individual terms by a factor 4. For the
topology \epsfig{figure=ladder.eps,height=2ex} we thus end up with 9
terms.

One can further see that the coefficients of the middle variables in
\(P_i\) are generated by subsequent multiplication of terms linear in
the edge variables (the \(\beta_{ij}\) in~(\ref{LemmaTwoSystem}) are
linear in edge variables according to~(\ref{shifteffect})) and can
therefore be factorized when inserted in the Heaviside functions.
Therefore the domains of integration in the edge variables are always
bounded by a number of linear hypersurfaces.

\begin{figure}[htb]
\epsfig{file=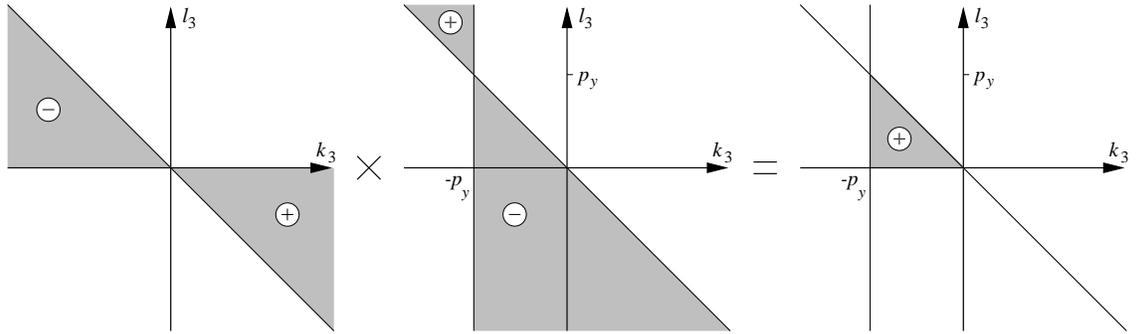,width=\textwidth}
\caption{Constraints after the \(l_2\)- and the \(k_2\)-integrations}
\label{fig:bounds1}
\end{figure}

E.g.\ when we perform the \(l_2\)-integration for a
scattering-amplitude of type \epsfig{figure=ladder.eps,height=2ex} and
apply lemma 1 on all the resulting residues, we obtain the infinite
domains sketched on the left of figure~\ref{fig:bounds1}. (The signs
denote the weight of the integrand there and stems from the numerator
in~(\ref{boxkorollar3})).  After the \(k_2\)-integration we obtain the
domain in the middle and when multiplied, only a finite triangle in
the \(k_3\)-\(l_3\)-plane remains.

Performing the \(l_1\)- and the \(k_1\)-integrations in the same way
we encounter poles on the real axis. They are, however, purely
artificial and a consequence of choosing equal imaginary parts
\(\eta\) in all propagators. Treating the integral as a Cauchy
Principal Value integral, the residue theorem can still be applied in
these cases since there are only odd coefficients in the integrand's
Laurent-series but the residue contributes only with a weight \(\pi
i\) instead of \(2\pi i\)~\cite{AblowitzFokas}.

\begin{figure}[tb]
\epsfig{file=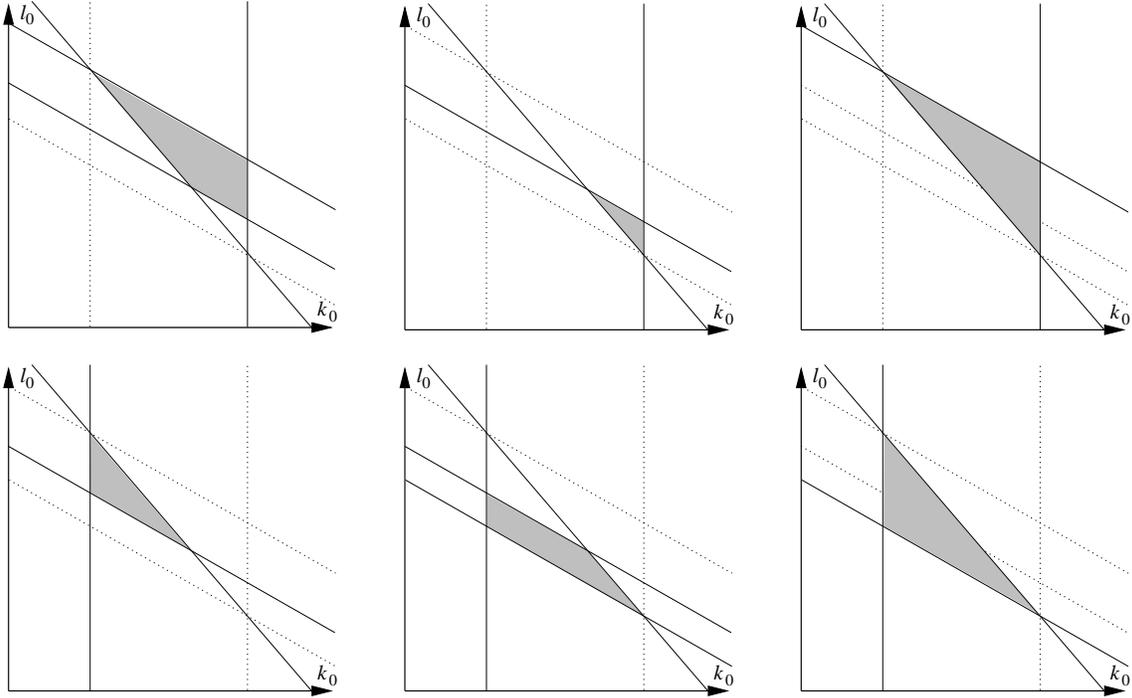,width=\textwidth}
\caption{Constraints after the \(l_1\)- and the \(k_1\)-integrations in topology I\,b). The exact parameters of the borders are functions of \(k_3\) and \(l_3\).}
\label{fig:bounds2}
\end{figure}

When this procedure is continued and all the constraints are combined
applying both lemmata, we get the 6 domains of integration in the
\(k_0\)-\(l_0\)-plane shown in figure~\ref{fig:bounds2} for a
scattering-amplitude \epsfig{figure=ladder.eps,height=2ex}. (There are
only 6 domains remaining instead of 9 due to an incompatibility of
constraints after the \(l_2\)- and the \(k_2\)-integrations in 3 of
them.) The exact parameters of the borders of these domains depend on
\(k_3\) and \(l_3\). We emphasize again, that the resulting domains of
integration in the edge variables are always finite and bounded by
linear functions---even in the case of crossed topologies---thus
making them accessible to an unambiguous reparametrization and
subsequent next integration. In the case of crossed topologies, lemma
2 can be applicable more than twice because of the presence of two
mixed propagators. The numbers of remaining terms turn out to be 12 in
\epsfig{figure=crossedladder.eps,height=2ex} and 5 in
\epsfig{figure=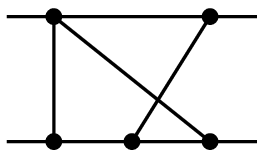,height=2ex}.

\section{Towards a 3-dimensional Integral}

The procedure outlined above produces a 4-fold integration of rational
function over a finite volume in the edge variables:
\begin{equation} \label{intdarst}
\int_{l_3^{(u)}}^{l_3^{(o)}}\!\!dl_3\int_{k_3^{(u)}}^{k_3^{(o)}}\!\!dk_3\int_{l_0^{(u)}}^{l_0^{(o)}}\!\!dl_0\int_{k_0^{(u)}}^{k_0^{(o)}}\!\!dk_0\,\frac{P(l_0,k_0,l_3,k_3)}{Q(l_0,k_0,l_3,k_3)}.
\end{equation}
The appearance of a numerator containing integration variables is due
to the solutions of \(P=0\) with respect to middle variables inserted
into the remaining propagators. After partial-fractioning this
integrand can always be transformed into one, where \(Q\) is no more
than quadratic in the edge variables. The integration-domains in the
\(k_0\)-\(l_0\)-plane which depend on \(k_3\) and \(l_3\) can be
mapped into some fixed domain independent of \(k_3\) and \(l_3\) using
a suitable linear transformation. The rational function can then be
integrated once more, resulting in other rational functions,
logarithms and arcustangens where care has to be taken of the small
imaginary part and the position of branch-cuts. The coefficients of
the result will generally involve square roots of quartic functions in
the remaining three variables. This three-dimensional integration
should be accessible to numerical evaluation using
\texttt{vegas}~\cite{Lepage78, Lepage80, PVegas} or similar routines.

\section{An Example}
\label{sec:example}

\begin{figure}[htb]
\begin{center}
\epsfig{file=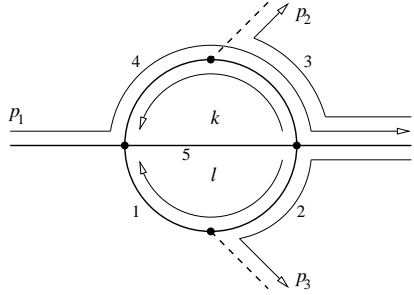}
\caption{Choice of momentum-flow for a decay-process of type III\,b)}
\end{center}
\label{fig:acnodeeggp}
\end{figure}

We will now test our method on a simple example: a decay-amplitude of
type \epsfig{figure=acnode.eps,height=2ex} where a heavy scalar
particle decays into a lighter and two massless ones as shown in
figure~\ref{fig:acnodeeggp}. The choice of momentum-flow indicated
there results in the inverse propagators
\begin{eqnarray}
  P_{l,1} = l^2 - m_1^2 + i\eta, & \quad & P_{l,2} = (l+p_3)^2 - m_2^2 + i\eta, \nonumber\\
  P_{k,3} = (k-p_1+p_2)^2 - m_3^2 + i\eta, & \quad & P_{k,4} = (k-p_1)^2 - m_4^2 + i\eta, \nonumber\\
  P_{l+k,5} = (l+k)^2 - m_5^2 + i\eta. & &
\end{eqnarray}

The most convenient Lorentz frame for this case turns out not to be
the rest-frame of the decaying particle but the one in which the
3-momenta of the light-like particles are antiparallel and aligned to
a specific coordinate-axis, say \(x\).  This frame can always be
reached by a Lorentz transformation as long as the two light-like
particles do not have equal momenta.

When integrating out the middle variables with these conditions and
applying the rules found in section~\ref{sec:constraints} one finds,
however, incompatible constraints in the \(l_1\)- and the
\(k_1\)-integrations. To get out of the dilemma, the light-cone
condition for the two massless external particles must be temporarily
relaxed\footnote{This difficulty is a manifestation of a degeneracy
  pointed out in~\cite{hepph9407234} which prohibits the application
  of our method to two- and three-point functions.}. A possible choice
of momenta is
\begin{eqnarray} \label{ExtMomenta}
 p_1^\mu=(q_\eta,p_x,p_y,0),\quad p_2^\mu=(q_\gamma,q_x,0,0),\quad p_3^\mu=(q_\gamma,-q_x,0,0), \nonumber\\
 p_4^\mu=p_1^\mu-p_2^\mu-p_3^\mu=(q_\eta-2\,q_\gamma,p_x,p_y,0),
\end{eqnarray}
where \(q_\gamma = \lim_{\epsilon \rightarrow +0} q_x+\epsilon\) is
tacitly assumed. 

\begin{figure}[htb]
\epsfig{file=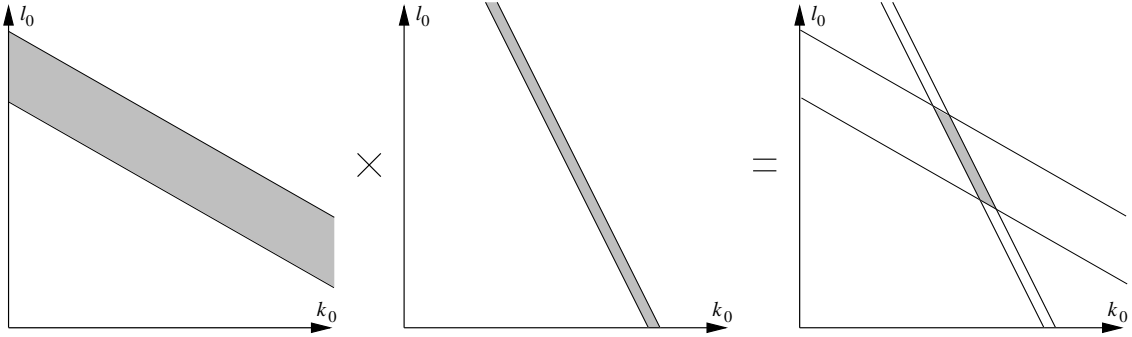,width=\textwidth}
\caption{Constraints after the \(l_1\)- and the \(k_1\)-integrations in topology III\,b)}
\label{fig:bounds3}
\end{figure}

Now the integrations in the middle variables can be performed
resulting in the domain of figure~\ref{fig:bounds3} for the \(k_0\)-
and the \(l_0\)-integrations in addition to the one already known from
figure~\ref{fig:bounds1} for the \(k_3\)- and \(l_3\)-integrations.

If we use a linear transformation in the \(k_0\)-\(l_0\)-plane in
order to map this domain into a unit-square \(\tilde{k}_0=(0\dots
1)\), \(\tilde{l}_0=(0\dots 1)\), another one to map the triangle in
the \(k_3\)-\(l_3\)-plane into a unit-triangle \(\tilde{k}_3=(0\dots
1)\), \(\tilde{l}_3=(0\dots 1-\tilde{k}_3)\), the limit
\(\epsilon\rightarrow 0\) can safely be performed and we obtain the
representation
\begin{equation}
 V(p_1,p_2,p_3) = {}-(2\pi i)^4\int_0^1\!\!\!d\tilde{k}_3\int_0^{1-\tilde{k}_3}\!\!\!\!d\tilde{l}_3\,\int_0^1\!\!\!d\tilde{k}_0\int_0^1\!\!\!d\tilde{l}_0\,\frac{1}{(b_1\,\tilde{k}_0+b_0)\,\tilde{l}_0+(a_1\,\tilde{k}_0+a_0)+\,i\eta}
\end{equation}
where the coefficients are given by
\begin{eqnarray}
 a_0 & = & 16\,\left(\tilde{k}_3\,\tilde{l}_3\,m_5^2\big/(\tilde{l}_3+\tilde{k}_3-1) - \tilde{k}_3\,m_1^2 - \tilde{l}_3\,m_3^2\right. \nonumber\\
  & & \left.\quad{}- \tilde{k}_3\,\tilde{l}_3\,\bigl(p_x^2+p_y^2-q_\eta^2+2\,q_\gamma\,(q_\eta-p_x)\bigr)\right), \nonumber\\
 a_1 & = & 16\,\tilde{l}_3\,\bigl(2\,q_\gamma\,\tilde{k}_3\,(q_\eta-p_x)+m_3^2-m_4^2\bigr), \\
 b_0 & = & 16\,\tilde{k}_3\,\bigl(2\,q_\gamma\,\tilde{l}_3\,(2\,q_\gamma-q_\eta-p_x)+m_1^2-m_2^2\bigr), \nonumber\\
 b_1 & = & -64\,\tilde{k}_3\,\tilde{l}_3\,q_\gamma^2. \nonumber
\end{eqnarray}
In this case, all terms quadratic in \(\tilde{l}_0\) or
\(\tilde{k}_0\) in the denominator have canceled, allowing for an
integration in both of these variables. Splitting the integral into
Principal Value integral and \(\delta\)-function separates real- and
imaginary part. We obtain a numerically stable two-fold representation
with dilogarithms in the real part and logarithms in the imaginary
part which displays the correct behaviour even at thresholds.

By letting \(q_\gamma\rightarrow 0\) we restrict our kinematical
regime even further and obtain a two-point sunset-topology with two
propagators squared. Numerical stability, however, breaks down as we
approach this limit. A look at the arguments of the dilogarithms in
our two-dimensional representation expanded in small \(q_\gamma\)
reveals why:
\begin{eqnarray} \label{sunsetexpand}
 \lefteqn{\lim_{q_\gamma\rightarrow0}\Re\big(V(p_1,p_2,p_3)\big)=\Re\big(P(p_1^2)\big)=-(2\pi i)^4\!\int_{0}^1\!\!\!d\tilde{k}_3\int_0^{1-\tilde{k}_3}\!\!\!\!d\tilde{l}_3} \nonumber\\
 & & \frac{1}{64\,\tilde{k}_3\,\tilde{l}_3\,q_\gamma^2}\;\Re\left(\li(1+\hat{r}_{24}\,q_\gamma^2)-\li(1+\hat{r}_{23}\,q_\gamma^2)-\li(1+\hat{r}_{14}\,q_\gamma^2)+\li(1+\hat{r}_{13}\,q_\gamma^2)\right).
\end{eqnarray}
The coefficients in this formula are
\begin{eqnarray} \label{T2entwickl}
\hat r_{ij} & = & \frac{4\,\tilde{k}_3\,\tilde{l}_3\,p_1^2}{(m_3^2-m_4^2)(m_1^2-m_2^2)} \nonumber\\
 & & {}+ 4\,\frac{(\tilde{k}_3\,m_i^2+\tilde{l}_3\,m_j^2)\,(1-\tilde{k}-\tilde{l})\,+\,m_5^2\,\tilde{l}_3\,\tilde{k}_3}{(m_3^2-m_4^2)(m_1^2-m_2^2)(\tilde{k}_3+\tilde{l}_3-1)}.
\end{eqnarray}
with \(p_1^2 = q_\eta^2-p_x^2-p_y^2\) as the only combination of
exterior momenta (c.f.\ eq.~(\ref{ExtMomenta})) entering \(P\), as it
should be.

Numerical stability can be restored in~(\ref{sunsetexpand}) if the
expansion of the dilogarithms at their branch-point as generalized
Taylor series
\begin{displaymath}
\Re\left(\li(1+\epsilon)\right)=\frac{\pi^2}{6} + \epsilon(1-\ln|\epsilon|)+O(\epsilon^2)
\end{displaymath}
and the relation
\(\hat{r}_{24}-\hat{r}_{23}-\hat{r}_{14}+\hat{r}_{13}=0\) are used.

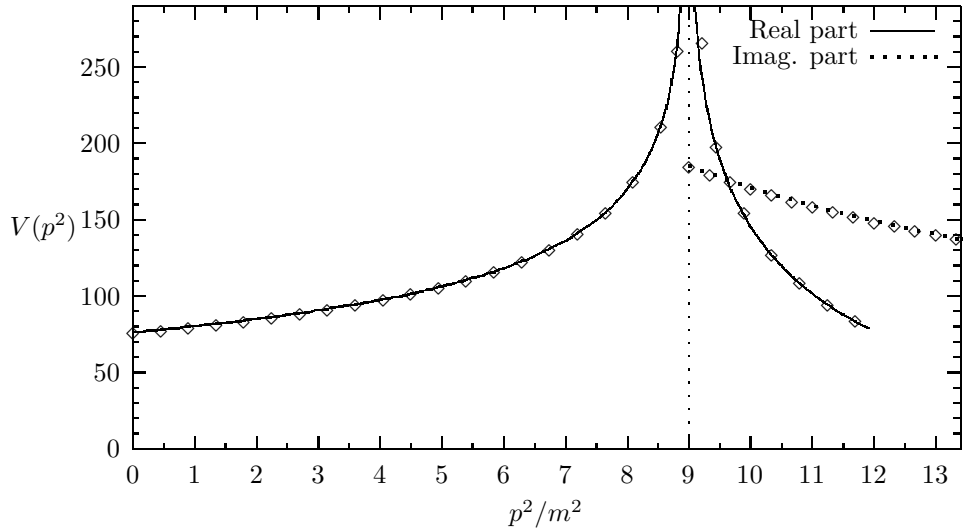
\begin{figure}[htb]
 \begin{center}
\setlength{\unitlength}{0.240900pt}
\ifx\plotpoint\undefined\newsavebox{\plotpoint}\fi
\sbox{\plotpoint}{\rule[-0.200pt]{0.400pt}{0.400pt}}%
\begin{picture}(1500,900)(0,0)
\font\gnuplot=cmr10 at 10pt
\gnuplot
\sbox{\plotpoint}{\rule[-0.200pt]{0.400pt}{0.400pt}}%
\put(181.0,163.0){\rule[-0.200pt]{4.818pt}{0.400pt}}
\put(161,163){\makebox(0,0)[r]{0}}
\put(1460.0,163.0){\rule[-0.200pt]{4.818pt}{0.400pt}}
\put(181.0,187.0){\rule[-0.200pt]{2.409pt}{0.400pt}}
\put(1470.0,187.0){\rule[-0.200pt]{2.409pt}{0.400pt}}
\put(181.0,211.0){\rule[-0.200pt]{2.409pt}{0.400pt}}
\put(1470.0,211.0){\rule[-0.200pt]{2.409pt}{0.400pt}}
\put(181.0,235.0){\rule[-0.200pt]{2.409pt}{0.400pt}}
\put(1470.0,235.0){\rule[-0.200pt]{2.409pt}{0.400pt}}
\put(181.0,259.0){\rule[-0.200pt]{2.409pt}{0.400pt}}
\put(1470.0,259.0){\rule[-0.200pt]{2.409pt}{0.400pt}}
\put(181.0,283.0){\rule[-0.200pt]{4.818pt}{0.400pt}}
\put(161,283){\makebox(0,0)[r]{50}}
\put(1460.0,283.0){\rule[-0.200pt]{4.818pt}{0.400pt}}
\put(181.0,307.0){\rule[-0.200pt]{2.409pt}{0.400pt}}
\put(1470.0,307.0){\rule[-0.200pt]{2.409pt}{0.400pt}}
\put(181.0,331.0){\rule[-0.200pt]{2.409pt}{0.400pt}}
\put(1470.0,331.0){\rule[-0.200pt]{2.409pt}{0.400pt}}
\put(181.0,355.0){\rule[-0.200pt]{2.409pt}{0.400pt}}
\put(1470.0,355.0){\rule[-0.200pt]{2.409pt}{0.400pt}}
\put(181.0,379.0){\rule[-0.200pt]{2.409pt}{0.400pt}}
\put(1470.0,379.0){\rule[-0.200pt]{2.409pt}{0.400pt}}
\put(181.0,403.0){\rule[-0.200pt]{4.818pt}{0.400pt}}
\put(161,403){\makebox(0,0)[r]{100}}
\put(1460.0,403.0){\rule[-0.200pt]{4.818pt}{0.400pt}}
\put(181.0,427.0){\rule[-0.200pt]{2.409pt}{0.400pt}}
\put(1470.0,427.0){\rule[-0.200pt]{2.409pt}{0.400pt}}
\put(181.0,451.0){\rule[-0.200pt]{2.409pt}{0.400pt}}
\put(1470.0,451.0){\rule[-0.200pt]{2.409pt}{0.400pt}}
\put(181.0,475.0){\rule[-0.200pt]{2.409pt}{0.400pt}}
\put(1470.0,475.0){\rule[-0.200pt]{2.409pt}{0.400pt}}
\put(181.0,499.0){\rule[-0.200pt]{2.409pt}{0.400pt}}
\put(1470.0,499.0){\rule[-0.200pt]{2.409pt}{0.400pt}}
\put(181.0,523.0){\rule[-0.200pt]{4.818pt}{0.400pt}}
\put(161,523){\makebox(0,0)[r]{150}}
\put(1460.0,523.0){\rule[-0.200pt]{4.818pt}{0.400pt}}
\put(181.0,547.0){\rule[-0.200pt]{2.409pt}{0.400pt}}
\put(1470.0,547.0){\rule[-0.200pt]{2.409pt}{0.400pt}}
\put(181.0,571.0){\rule[-0.200pt]{2.409pt}{0.400pt}}
\put(1470.0,571.0){\rule[-0.200pt]{2.409pt}{0.400pt}}
\put(181.0,595.0){\rule[-0.200pt]{2.409pt}{0.400pt}}
\put(1470.0,595.0){\rule[-0.200pt]{2.409pt}{0.400pt}}
\put(181.0,619.0){\rule[-0.200pt]{2.409pt}{0.400pt}}
\put(1470.0,619.0){\rule[-0.200pt]{2.409pt}{0.400pt}}
\put(181.0,643.0){\rule[-0.200pt]{4.818pt}{0.400pt}}
\put(161,643){\makebox(0,0)[r]{200}}
\put(1460.0,643.0){\rule[-0.200pt]{4.818pt}{0.400pt}}
\put(181.0,667.0){\rule[-0.200pt]{2.409pt}{0.400pt}}
\put(1470.0,667.0){\rule[-0.200pt]{2.409pt}{0.400pt}}
\put(181.0,691.0){\rule[-0.200pt]{2.409pt}{0.400pt}}
\put(1470.0,691.0){\rule[-0.200pt]{2.409pt}{0.400pt}}
\put(181.0,715.0){\rule[-0.200pt]{2.409pt}{0.400pt}}
\put(1470.0,715.0){\rule[-0.200pt]{2.409pt}{0.400pt}}
\put(181.0,739.0){\rule[-0.200pt]{2.409pt}{0.400pt}}
\put(1470.0,739.0){\rule[-0.200pt]{2.409pt}{0.400pt}}
\put(181.0,763.0){\rule[-0.200pt]{4.818pt}{0.400pt}}
\put(161,763){\makebox(0,0)[r]{250}}
\put(1460.0,763.0){\rule[-0.200pt]{4.818pt}{0.400pt}}
\put(181.0,787.0){\rule[-0.200pt]{2.409pt}{0.400pt}}
\put(1470.0,787.0){\rule[-0.200pt]{2.409pt}{0.400pt}}
\put(181.0,811.0){\rule[-0.200pt]{2.409pt}{0.400pt}}
\put(1470.0,811.0){\rule[-0.200pt]{2.409pt}{0.400pt}}
\put(181.0,835.0){\rule[-0.200pt]{2.409pt}{0.400pt}}
\put(1470.0,835.0){\rule[-0.200pt]{2.409pt}{0.400pt}}
\put(181.0,859.0){\rule[-0.200pt]{2.409pt}{0.400pt}}
\put(1470.0,859.0){\rule[-0.200pt]{2.409pt}{0.400pt}}
\put(181.0,163.0){\rule[-0.200pt]{0.400pt}{4.818pt}}
\put(181,122){\makebox(0,0){0}}
\put(181.0,839.0){\rule[-0.200pt]{0.400pt}{4.818pt}}
\put(229.0,163.0){\rule[-0.200pt]{0.400pt}{2.409pt}}
\put(229.0,849.0){\rule[-0.200pt]{0.400pt}{2.409pt}}
\put(278.0,163.0){\rule[-0.200pt]{0.400pt}{4.818pt}}
\put(278,122){\makebox(0,0){1}}
\put(278.0,839.0){\rule[-0.200pt]{0.400pt}{4.818pt}}
\put(326.0,163.0){\rule[-0.200pt]{0.400pt}{2.409pt}}
\put(326.0,849.0){\rule[-0.200pt]{0.400pt}{2.409pt}}
\put(375.0,163.0){\rule[-0.200pt]{0.400pt}{4.818pt}}
\put(375,122){\makebox(0,0){2}}
\put(375.0,839.0){\rule[-0.200pt]{0.400pt}{4.818pt}}
\put(423.0,163.0){\rule[-0.200pt]{0.400pt}{2.409pt}}
\put(423.0,849.0){\rule[-0.200pt]{0.400pt}{2.409pt}}
\put(472.0,163.0){\rule[-0.200pt]{0.400pt}{4.818pt}}
\put(472,122){\makebox(0,0){3}}
\put(472.0,839.0){\rule[-0.200pt]{0.400pt}{4.818pt}}
\put(520.0,163.0){\rule[-0.200pt]{0.400pt}{2.409pt}}
\put(520.0,849.0){\rule[-0.200pt]{0.400pt}{2.409pt}}
\put(569.0,163.0){\rule[-0.200pt]{0.400pt}{4.818pt}}
\put(569,122){\makebox(0,0){4}}
\put(569.0,839.0){\rule[-0.200pt]{0.400pt}{4.818pt}}
\put(617.0,163.0){\rule[-0.200pt]{0.400pt}{2.409pt}}
\put(617.0,849.0){\rule[-0.200pt]{0.400pt}{2.409pt}}
\put(666.0,163.0){\rule[-0.200pt]{0.400pt}{4.818pt}}
\put(666,122){\makebox(0,0){5}}
\put(666.0,839.0){\rule[-0.200pt]{0.400pt}{4.818pt}}
\put(714.0,163.0){\rule[-0.200pt]{0.400pt}{2.409pt}}
\put(714.0,849.0){\rule[-0.200pt]{0.400pt}{2.409pt}}
\put(763.0,163.0){\rule[-0.200pt]{0.400pt}{4.818pt}}
\put(763,122){\makebox(0,0){6}}
\put(763.0,839.0){\rule[-0.200pt]{0.400pt}{4.818pt}}
\put(811.0,163.0){\rule[-0.200pt]{0.400pt}{2.409pt}}
\put(811.0,849.0){\rule[-0.200pt]{0.400pt}{2.409pt}}
\put(860.0,163.0){\rule[-0.200pt]{0.400pt}{4.818pt}}
\put(860,122){\makebox(0,0){7}}
\put(860.0,839.0){\rule[-0.200pt]{0.400pt}{4.818pt}}
\put(908.0,163.0){\rule[-0.200pt]{0.400pt}{2.409pt}}
\put(908.0,849.0){\rule[-0.200pt]{0.400pt}{2.409pt}}
\put(957.0,163.0){\rule[-0.200pt]{0.400pt}{4.818pt}}
\put(957,122){\makebox(0,0){8}}
\put(957.0,839.0){\rule[-0.200pt]{0.400pt}{4.818pt}}
\put(1005.0,163.0){\rule[-0.200pt]{0.400pt}{2.409pt}}
\put(1005.0,849.0){\rule[-0.200pt]{0.400pt}{2.409pt}}
\put(1053.0,163.0){\rule[-0.200pt]{0.400pt}{4.818pt}}
\put(1053,122){\makebox(0,0){9}}
\put(1053.0,839.0){\rule[-0.200pt]{0.400pt}{4.818pt}}
\put(1102.0,163.0){\rule[-0.200pt]{0.400pt}{2.409pt}}
\put(1102.0,849.0){\rule[-0.200pt]{0.400pt}{2.409pt}}
\put(1150.0,163.0){\rule[-0.200pt]{0.400pt}{4.818pt}}
\put(1150,122){\makebox(0,0){10}}
\put(1150.0,839.0){\rule[-0.200pt]{0.400pt}{4.818pt}}
\put(1199.0,163.0){\rule[-0.200pt]{0.400pt}{2.409pt}}
\put(1199.0,849.0){\rule[-0.200pt]{0.400pt}{2.409pt}}
\put(1247.0,163.0){\rule[-0.200pt]{0.400pt}{4.818pt}}
\put(1247,122){\makebox(0,0){11}}
\put(1247.0,839.0){\rule[-0.200pt]{0.400pt}{4.818pt}}
\put(1296.0,163.0){\rule[-0.200pt]{0.400pt}{2.409pt}}
\put(1296.0,849.0){\rule[-0.200pt]{0.400pt}{2.409pt}}
\put(1344.0,163.0){\rule[-0.200pt]{0.400pt}{4.818pt}}
\put(1344,122){\makebox(0,0){12}}
\put(1344.0,839.0){\rule[-0.200pt]{0.400pt}{4.818pt}}
\put(1393.0,163.0){\rule[-0.200pt]{0.400pt}{2.409pt}}
\put(1393.0,849.0){\rule[-0.200pt]{0.400pt}{2.409pt}}
\put(1441.0,163.0){\rule[-0.200pt]{0.400pt}{4.818pt}}
\put(1441,122){\makebox(0,0){13}}
\put(1441.0,839.0){\rule[-0.200pt]{0.400pt}{4.818pt}}
\put(181.0,163.0){\rule[-0.200pt]{312.929pt}{0.400pt}}
\put(1480.0,163.0){\rule[-0.200pt]{0.400pt}{167.666pt}}
\put(181.0,859.0){\rule[-0.200pt]{312.929pt}{0.400pt}}
\put(41,511){\makebox(0,0){$V(p^2)$}}
\put(830,61){\makebox(0,0){$p^2/m^2$}}
\put(181.0,163.0){\rule[-0.200pt]{0.400pt}{167.666pt}}
\put(1320,819){\makebox(0,0)[r]{Real part}}
\put(1340.0,819.0){\rule[-0.200pt]{24.090pt}{0.400pt}}
\put(1335,353){\usebox{\plotpoint}}
\multiput(1331.26,353.59)(-1.044,0.477){7}{\rule{0.900pt}{0.115pt}}
\multiput(1333.13,352.17)(-8.132,5.000){2}{\rule{0.450pt}{0.400pt}}
\multiput(1321.26,358.59)(-1.044,0.477){7}{\rule{0.900pt}{0.115pt}}
\multiput(1323.13,357.17)(-8.132,5.000){2}{\rule{0.450pt}{0.400pt}}
\multiput(1312.09,363.59)(-0.762,0.482){9}{\rule{0.700pt}{0.116pt}}
\multiput(1313.55,362.17)(-7.547,6.000){2}{\rule{0.350pt}{0.400pt}}
\multiput(1302.26,369.59)(-1.044,0.477){7}{\rule{0.900pt}{0.115pt}}
\multiput(1304.13,368.17)(-8.132,5.000){2}{\rule{0.450pt}{0.400pt}}
\multiput(1292.82,374.59)(-0.852,0.482){9}{\rule{0.767pt}{0.116pt}}
\multiput(1294.41,373.17)(-8.409,6.000){2}{\rule{0.383pt}{0.400pt}}
\multiput(1283.21,380.59)(-0.721,0.485){11}{\rule{0.671pt}{0.117pt}}
\multiput(1284.61,379.17)(-8.606,7.000){2}{\rule{0.336pt}{0.400pt}}
\multiput(1273.09,387.59)(-0.762,0.482){9}{\rule{0.700pt}{0.116pt}}
\multiput(1274.55,386.17)(-7.547,6.000){2}{\rule{0.350pt}{0.400pt}}
\multiput(1264.21,393.59)(-0.721,0.485){11}{\rule{0.671pt}{0.117pt}}
\multiput(1265.61,392.17)(-8.606,7.000){2}{\rule{0.336pt}{0.400pt}}
\multiput(1254.51,400.59)(-0.626,0.488){13}{\rule{0.600pt}{0.117pt}}
\multiput(1255.75,399.17)(-8.755,8.000){2}{\rule{0.300pt}{0.400pt}}
\multiput(1244.45,408.59)(-0.645,0.485){11}{\rule{0.614pt}{0.117pt}}
\multiput(1245.73,407.17)(-7.725,7.000){2}{\rule{0.307pt}{0.400pt}}
\multiput(1235.51,415.59)(-0.626,0.488){13}{\rule{0.600pt}{0.117pt}}
\multiput(1236.75,414.17)(-8.755,8.000){2}{\rule{0.300pt}{0.400pt}}
\multiput(1225.74,423.59)(-0.553,0.489){15}{\rule{0.544pt}{0.118pt}}
\multiput(1226.87,422.17)(-8.870,9.000){2}{\rule{0.272pt}{0.400pt}}
\multiput(1215.92,432.59)(-0.495,0.489){15}{\rule{0.500pt}{0.118pt}}
\multiput(1216.96,431.17)(-7.962,9.000){2}{\rule{0.250pt}{0.400pt}}
\multiput(1206.92,441.58)(-0.495,0.491){17}{\rule{0.500pt}{0.118pt}}
\multiput(1207.96,440.17)(-8.962,10.000){2}{\rule{0.250pt}{0.400pt}}
\multiput(1197.92,451.00)(-0.491,0.547){17}{\rule{0.118pt}{0.540pt}}
\multiput(1198.17,451.00)(-10.000,9.879){2}{\rule{0.400pt}{0.270pt}}
\multiput(1187.92,462.00)(-0.491,0.547){17}{\rule{0.118pt}{0.540pt}}
\multiput(1188.17,462.00)(-10.000,9.879){2}{\rule{0.400pt}{0.270pt}}
\multiput(1177.93,473.00)(-0.489,0.669){15}{\rule{0.118pt}{0.633pt}}
\multiput(1178.17,473.00)(-9.000,10.685){2}{\rule{0.400pt}{0.317pt}}
\multiput(1168.92,485.00)(-0.491,0.704){17}{\rule{0.118pt}{0.660pt}}
\multiput(1169.17,485.00)(-10.000,12.630){2}{\rule{0.400pt}{0.330pt}}
\multiput(1158.92,499.00)(-0.491,0.704){17}{\rule{0.118pt}{0.660pt}}
\multiput(1159.17,499.00)(-10.000,12.630){2}{\rule{0.400pt}{0.330pt}}
\multiput(1148.93,513.00)(-0.489,0.961){15}{\rule{0.118pt}{0.856pt}}
\multiput(1149.17,513.00)(-9.000,15.224){2}{\rule{0.400pt}{0.428pt}}
\multiput(1139.92,530.00)(-0.491,0.912){17}{\rule{0.118pt}{0.820pt}}
\multiput(1140.17,530.00)(-10.000,16.298){2}{\rule{0.400pt}{0.410pt}}
\multiput(1129.92,548.00)(-0.491,1.017){17}{\rule{0.118pt}{0.900pt}}
\multiput(1130.17,548.00)(-10.000,18.132){2}{\rule{0.400pt}{0.450pt}}
\multiput(1119.93,568.00)(-0.489,1.368){15}{\rule{0.118pt}{1.167pt}}
\multiput(1120.17,568.00)(-9.000,21.579){2}{\rule{0.400pt}{0.583pt}}
\multiput(1110.92,592.00)(-0.491,1.433){17}{\rule{0.118pt}{1.220pt}}
\multiput(1111.17,592.00)(-10.000,25.468){2}{\rule{0.400pt}{0.610pt}}
\multiput(1100.92,620.00)(-0.491,1.746){17}{\rule{0.118pt}{1.460pt}}
\multiput(1101.17,620.00)(-10.000,30.970){2}{\rule{0.400pt}{0.730pt}}
\multiput(1090.93,654.00)(-0.489,2.475){15}{\rule{0.118pt}{2.011pt}}
\multiput(1091.17,654.00)(-9.000,38.826){2}{\rule{0.400pt}{1.006pt}}
\multiput(1081.92,697.00)(-0.491,3.101){17}{\rule{0.118pt}{2.500pt}}
\multiput(1082.17,697.00)(-10.000,54.811){2}{\rule{0.400pt}{1.250pt}}
\multiput(1071.92,757.00)(-0.491,5.291){17}{\rule{0.118pt}{4.180pt}}
\multiput(1072.17,757.00)(-10.000,93.324){2}{\rule{0.400pt}{2.090pt}}
\multiput(1039.93,837.00)(-0.477,-7.167){7}{\rule{0.115pt}{5.300pt}}
\multiput(1040.17,848.00)(-5.000,-54.000){2}{\rule{0.400pt}{2.650pt}}
\multiput(1034.93,783.44)(-0.489,-3.174){15}{\rule{0.118pt}{2.544pt}}
\multiput(1035.17,788.72)(-9.000,-49.719){2}{\rule{0.400pt}{1.272pt}}
\multiput(1025.93,730.28)(-0.488,-2.607){13}{\rule{0.117pt}{2.100pt}}
\multiput(1026.17,734.64)(-8.000,-35.641){2}{\rule{0.400pt}{1.050pt}}
\multiput(1017.93,693.42)(-0.489,-1.601){15}{\rule{0.118pt}{1.344pt}}
\multiput(1018.17,696.21)(-9.000,-25.210){2}{\rule{0.400pt}{0.672pt}}
\multiput(1008.93,665.97)(-0.489,-1.427){15}{\rule{0.118pt}{1.211pt}}
\multiput(1009.17,668.49)(-9.000,-22.486){2}{\rule{0.400pt}{0.606pt}}
\multiput(999.93,641.90)(-0.489,-1.135){15}{\rule{0.118pt}{0.989pt}}
\multiput(1000.17,643.95)(-9.000,-17.948){2}{\rule{0.400pt}{0.494pt}}
\multiput(990.93,622.26)(-0.488,-1.022){13}{\rule{0.117pt}{0.900pt}}
\multiput(991.17,624.13)(-8.000,-14.132){2}{\rule{0.400pt}{0.450pt}}
\multiput(982.93,606.82)(-0.489,-0.844){15}{\rule{0.118pt}{0.767pt}}
\multiput(983.17,608.41)(-9.000,-13.409){2}{\rule{0.400pt}{0.383pt}}
\multiput(973.93,592.56)(-0.489,-0.611){15}{\rule{0.118pt}{0.589pt}}
\multiput(974.17,593.78)(-9.000,-9.778){2}{\rule{0.400pt}{0.294pt}}
\multiput(964.93,581.19)(-0.489,-0.728){15}{\rule{0.118pt}{0.678pt}}
\multiput(965.17,582.59)(-9.000,-11.593){2}{\rule{0.400pt}{0.339pt}}
\multiput(955.93,568.30)(-0.488,-0.692){13}{\rule{0.117pt}{0.650pt}}
\multiput(956.17,569.65)(-8.000,-9.651){2}{\rule{0.400pt}{0.325pt}}
\multiput(946.92,558.93)(-0.495,-0.489){15}{\rule{0.500pt}{0.118pt}}
\multiput(947.96,559.17)(-7.962,-9.000){2}{\rule{0.250pt}{0.400pt}}
\multiput(938.93,548.74)(-0.489,-0.553){15}{\rule{0.118pt}{0.544pt}}
\multiput(939.17,549.87)(-9.000,-8.870){2}{\rule{0.400pt}{0.272pt}}
\multiput(928.69,539.93)(-0.569,-0.485){11}{\rule{0.557pt}{0.117pt}}
\multiput(929.84,540.17)(-6.844,-7.000){2}{\rule{0.279pt}{0.400pt}}
\multiput(920.72,532.93)(-0.560,-0.488){13}{\rule{0.550pt}{0.117pt}}
\multiput(921.86,533.17)(-7.858,-8.000){2}{\rule{0.275pt}{0.400pt}}
\multiput(911.45,524.93)(-0.645,-0.485){11}{\rule{0.614pt}{0.117pt}}
\multiput(912.73,525.17)(-7.725,-7.000){2}{\rule{0.307pt}{0.400pt}}
\multiput(902.09,517.93)(-0.762,-0.482){9}{\rule{0.700pt}{0.116pt}}
\multiput(903.55,518.17)(-7.547,-6.000){2}{\rule{0.350pt}{0.400pt}}
\multiput(892.93,511.93)(-0.821,-0.477){7}{\rule{0.740pt}{0.115pt}}
\multiput(894.46,512.17)(-6.464,-5.000){2}{\rule{0.370pt}{0.400pt}}
\multiput(885.09,506.93)(-0.762,-0.482){9}{\rule{0.700pt}{0.116pt}}
\multiput(886.55,507.17)(-7.547,-6.000){2}{\rule{0.350pt}{0.400pt}}
\multiput(876.45,500.93)(-0.645,-0.485){11}{\rule{0.614pt}{0.117pt}}
\multiput(877.73,501.17)(-7.725,-7.000){2}{\rule{0.307pt}{0.400pt}}
\multiput(866.93,493.93)(-0.821,-0.477){7}{\rule{0.740pt}{0.115pt}}
\multiput(868.46,494.17)(-6.464,-5.000){2}{\rule{0.370pt}{0.400pt}}
\multiput(857.85,488.94)(-1.212,-0.468){5}{\rule{1.000pt}{0.113pt}}
\multiput(859.92,489.17)(-6.924,-4.000){2}{\rule{0.500pt}{0.400pt}}
\multiput(848.85,484.94)(-1.212,-0.468){5}{\rule{1.000pt}{0.113pt}}
\multiput(850.92,485.17)(-6.924,-4.000){2}{\rule{0.500pt}{0.400pt}}
\multiput(840.60,480.93)(-0.933,-0.477){7}{\rule{0.820pt}{0.115pt}}
\multiput(842.30,481.17)(-7.298,-5.000){2}{\rule{0.410pt}{0.400pt}}
\multiput(831.26,475.94)(-1.066,-0.468){5}{\rule{0.900pt}{0.113pt}}
\multiput(833.13,476.17)(-6.132,-4.000){2}{\rule{0.450pt}{0.400pt}}
\multiput(822.85,471.94)(-1.212,-0.468){5}{\rule{1.000pt}{0.113pt}}
\multiput(824.92,472.17)(-6.924,-4.000){2}{\rule{0.500pt}{0.400pt}}
\multiput(814.60,467.93)(-0.933,-0.477){7}{\rule{0.820pt}{0.115pt}}
\multiput(816.30,468.17)(-7.298,-5.000){2}{\rule{0.410pt}{0.400pt}}
\multiput(804.85,462.94)(-1.212,-0.468){5}{\rule{1.000pt}{0.113pt}}
\multiput(806.92,463.17)(-6.924,-4.000){2}{\rule{0.500pt}{0.400pt}}
\put(792,458.17){\rule{1.700pt}{0.400pt}}
\multiput(796.47,459.17)(-4.472,-2.000){2}{\rule{0.850pt}{0.400pt}}
\multiput(787.85,456.94)(-1.212,-0.468){5}{\rule{1.000pt}{0.113pt}}
\multiput(789.92,457.17)(-6.924,-4.000){2}{\rule{0.500pt}{0.400pt}}
\multiput(778.85,452.94)(-1.212,-0.468){5}{\rule{1.000pt}{0.113pt}}
\multiput(780.92,453.17)(-6.924,-4.000){2}{\rule{0.500pt}{0.400pt}}
\multiput(769.16,448.95)(-1.579,-0.447){3}{\rule{1.167pt}{0.108pt}}
\multiput(771.58,449.17)(-5.579,-3.000){2}{\rule{0.583pt}{0.400pt}}
\put(757,445.17){\rule{1.900pt}{0.400pt}}
\multiput(762.06,446.17)(-5.056,-2.000){2}{\rule{0.950pt}{0.400pt}}
\multiput(753.60,443.93)(-0.933,-0.477){7}{\rule{0.820pt}{0.115pt}}
\multiput(755.30,444.17)(-7.298,-5.000){2}{\rule{0.410pt}{0.400pt}}
\put(739,438.67){\rule{2.168pt}{0.400pt}}
\multiput(743.50,439.17)(-4.500,-1.000){2}{\rule{1.084pt}{0.400pt}}
\multiput(735.26,437.94)(-1.066,-0.468){5}{\rule{0.900pt}{0.113pt}}
\multiput(737.13,438.17)(-6.132,-4.000){2}{\rule{0.450pt}{0.400pt}}
\put(722,433.17){\rule{1.900pt}{0.400pt}}
\multiput(727.06,434.17)(-5.056,-2.000){2}{\rule{0.950pt}{0.400pt}}
\put(713,431.17){\rule{1.900pt}{0.400pt}}
\multiput(718.06,432.17)(-5.056,-2.000){2}{\rule{0.950pt}{0.400pt}}
\multiput(707.60,429.95)(-1.802,-0.447){3}{\rule{1.300pt}{0.108pt}}
\multiput(710.30,430.17)(-6.302,-3.000){2}{\rule{0.650pt}{0.400pt}}
\put(696,426.17){\rule{1.700pt}{0.400pt}}
\multiput(700.47,427.17)(-4.472,-2.000){2}{\rule{0.850pt}{0.400pt}}
\multiput(690.60,424.95)(-1.802,-0.447){3}{\rule{1.300pt}{0.108pt}}
\multiput(693.30,425.17)(-6.302,-3.000){2}{\rule{0.650pt}{0.400pt}}
\put(678,421.17){\rule{1.900pt}{0.400pt}}
\multiput(683.06,422.17)(-5.056,-2.000){2}{\rule{0.950pt}{0.400pt}}
\put(670,419.17){\rule{1.700pt}{0.400pt}}
\multiput(674.47,420.17)(-4.472,-2.000){2}{\rule{0.850pt}{0.400pt}}
\put(661,417.17){\rule{1.900pt}{0.400pt}}
\multiput(666.06,418.17)(-5.056,-2.000){2}{\rule{0.950pt}{0.400pt}}
\multiput(655.60,415.95)(-1.802,-0.447){3}{\rule{1.300pt}{0.108pt}}
\multiput(658.30,416.17)(-6.302,-3.000){2}{\rule{0.650pt}{0.400pt}}
\put(643,412.67){\rule{2.168pt}{0.400pt}}
\multiput(647.50,413.17)(-4.500,-1.000){2}{\rule{1.084pt}{0.400pt}}
\multiput(638.16,411.95)(-1.579,-0.447){3}{\rule{1.167pt}{0.108pt}}
\multiput(640.58,412.17)(-5.579,-3.000){2}{\rule{0.583pt}{0.400pt}}
\put(626,408.17){\rule{1.900pt}{0.400pt}}
\multiput(631.06,409.17)(-5.056,-2.000){2}{\rule{0.950pt}{0.400pt}}
\put(617,406.17){\rule{1.900pt}{0.400pt}}
\multiput(622.06,407.17)(-5.056,-2.000){2}{\rule{0.950pt}{0.400pt}}
\put(609,404.17){\rule{1.700pt}{0.400pt}}
\multiput(613.47,405.17)(-4.472,-2.000){2}{\rule{0.850pt}{0.400pt}}
\put(600,402.17){\rule{1.900pt}{0.400pt}}
\multiput(605.06,403.17)(-5.056,-2.000){2}{\rule{0.950pt}{0.400pt}}
\put(591,400.67){\rule{2.168pt}{0.400pt}}
\multiput(595.50,401.17)(-4.500,-1.000){2}{\rule{1.084pt}{0.400pt}}
\put(582,399.17){\rule{1.900pt}{0.400pt}}
\multiput(587.06,400.17)(-5.056,-2.000){2}{\rule{0.950pt}{0.400pt}}
\put(574,397.67){\rule{1.927pt}{0.400pt}}
\multiput(578.00,398.17)(-4.000,-1.000){2}{\rule{0.964pt}{0.400pt}}
\put(565,396.17){\rule{1.900pt}{0.400pt}}
\multiput(570.06,397.17)(-5.056,-2.000){2}{\rule{0.950pt}{0.400pt}}
\put(556,394.17){\rule{1.900pt}{0.400pt}}
\multiput(561.06,395.17)(-5.056,-2.000){2}{\rule{0.950pt}{0.400pt}}
\put(547,392.67){\rule{2.168pt}{0.400pt}}
\multiput(551.50,393.17)(-4.500,-1.000){2}{\rule{1.084pt}{0.400pt}}
\put(539,391.17){\rule{1.700pt}{0.400pt}}
\multiput(543.47,392.17)(-4.472,-2.000){2}{\rule{0.850pt}{0.400pt}}
\put(530,389.17){\rule{1.900pt}{0.400pt}}
\multiput(535.06,390.17)(-5.056,-2.000){2}{\rule{0.950pt}{0.400pt}}
\put(513,387.17){\rule{1.700pt}{0.400pt}}
\multiput(517.47,388.17)(-4.472,-2.000){2}{\rule{0.850pt}{0.400pt}}
\put(504,385.67){\rule{2.168pt}{0.400pt}}
\multiput(508.50,386.17)(-4.500,-1.000){2}{\rule{1.084pt}{0.400pt}}
\put(495,384.67){\rule{2.168pt}{0.400pt}}
\multiput(499.50,385.17)(-4.500,-1.000){2}{\rule{1.084pt}{0.400pt}}
\multiput(489.60,383.95)(-1.802,-0.447){3}{\rule{1.300pt}{0.108pt}}
\multiput(492.30,384.17)(-6.302,-3.000){2}{\rule{0.650pt}{0.400pt}}
\put(478,380.67){\rule{1.927pt}{0.400pt}}
\multiput(482.00,381.17)(-4.000,-1.000){2}{\rule{0.964pt}{0.400pt}}
\put(521.0,389.0){\rule[-0.200pt]{2.168pt}{0.400pt}}
\multiput(463.60,379.95)(-1.802,-0.447){3}{\rule{1.300pt}{0.108pt}}
\multiput(466.30,380.17)(-6.302,-3.000){2}{\rule{0.650pt}{0.400pt}}
\put(451,376.17){\rule{1.900pt}{0.400pt}}
\multiput(456.06,377.17)(-5.056,-2.000){2}{\rule{0.950pt}{0.400pt}}
\put(469.0,381.0){\rule[-0.200pt]{2.168pt}{0.400pt}}
\put(434,374.67){\rule{2.168pt}{0.400pt}}
\multiput(438.50,375.17)(-4.500,-1.000){2}{\rule{1.084pt}{0.400pt}}
\put(425,373.67){\rule{2.168pt}{0.400pt}}
\multiput(429.50,374.17)(-4.500,-1.000){2}{\rule{1.084pt}{0.400pt}}
\put(417,372.17){\rule{1.700pt}{0.400pt}}
\multiput(421.47,373.17)(-4.472,-2.000){2}{\rule{0.850pt}{0.400pt}}
\put(408,370.67){\rule{2.168pt}{0.400pt}}
\multiput(412.50,371.17)(-4.500,-1.000){2}{\rule{1.084pt}{0.400pt}}
\put(399,369.67){\rule{2.168pt}{0.400pt}}
\multiput(403.50,370.17)(-4.500,-1.000){2}{\rule{1.084pt}{0.400pt}}
\put(390,368.67){\rule{2.168pt}{0.400pt}}
\multiput(394.50,369.17)(-4.500,-1.000){2}{\rule{1.084pt}{0.400pt}}
\put(382,367.67){\rule{1.927pt}{0.400pt}}
\multiput(386.00,368.17)(-4.000,-1.000){2}{\rule{0.964pt}{0.400pt}}
\put(373,366.67){\rule{2.168pt}{0.400pt}}
\multiput(377.50,367.17)(-4.500,-1.000){2}{\rule{1.084pt}{0.400pt}}
\put(364,365.67){\rule{2.168pt}{0.400pt}}
\multiput(368.50,366.17)(-4.500,-1.000){2}{\rule{1.084pt}{0.400pt}}
\put(355,364.17){\rule{1.900pt}{0.400pt}}
\multiput(360.06,365.17)(-5.056,-2.000){2}{\rule{0.950pt}{0.400pt}}
\put(347,362.67){\rule{1.927pt}{0.400pt}}
\multiput(351.00,363.17)(-4.000,-1.000){2}{\rule{0.964pt}{0.400pt}}
\put(338,361.67){\rule{2.168pt}{0.400pt}}
\multiput(342.50,362.17)(-4.500,-1.000){2}{\rule{1.084pt}{0.400pt}}
\put(329,360.67){\rule{2.168pt}{0.400pt}}
\multiput(333.50,361.17)(-4.500,-1.000){2}{\rule{1.084pt}{0.400pt}}
\put(443.0,376.0){\rule[-0.200pt]{1.927pt}{0.400pt}}
\put(312,359.17){\rule{1.900pt}{0.400pt}}
\multiput(317.06,360.17)(-5.056,-2.000){2}{\rule{0.950pt}{0.400pt}}
\put(321.0,361.0){\rule[-0.200pt]{1.927pt}{0.400pt}}
\put(294,357.67){\rule{2.168pt}{0.400pt}}
\multiput(298.50,358.17)(-4.500,-1.000){2}{\rule{1.084pt}{0.400pt}}
\put(286,356.17){\rule{1.700pt}{0.400pt}}
\multiput(290.47,357.17)(-4.472,-2.000){2}{\rule{0.850pt}{0.400pt}}
\put(303.0,359.0){\rule[-0.200pt]{2.168pt}{0.400pt}}
\put(268,354.67){\rule{2.168pt}{0.400pt}}
\multiput(272.50,355.17)(-4.500,-1.000){2}{\rule{1.084pt}{0.400pt}}
\put(260,353.67){\rule{1.927pt}{0.400pt}}
\multiput(264.00,354.17)(-4.000,-1.000){2}{\rule{0.964pt}{0.400pt}}
\put(251,352.17){\rule{1.900pt}{0.400pt}}
\multiput(256.06,353.17)(-5.056,-2.000){2}{\rule{0.950pt}{0.400pt}}
\put(277.0,356.0){\rule[-0.200pt]{2.168pt}{0.400pt}}
\put(233,350.67){\rule{2.168pt}{0.400pt}}
\multiput(237.50,351.17)(-4.500,-1.000){2}{\rule{1.084pt}{0.400pt}}
\put(225,349.67){\rule{1.927pt}{0.400pt}}
\multiput(229.00,350.17)(-4.000,-1.000){2}{\rule{0.964pt}{0.400pt}}
\put(216,348.67){\rule{2.168pt}{0.400pt}}
\multiput(220.50,349.17)(-4.500,-1.000){2}{\rule{1.084pt}{0.400pt}}
\put(242.0,352.0){\rule[-0.200pt]{2.168pt}{0.400pt}}
\put(198,347.17){\rule{1.900pt}{0.400pt}}
\multiput(203.06,348.17)(-5.056,-2.000){2}{\rule{0.950pt}{0.400pt}}
\put(207.0,349.0){\rule[-0.200pt]{2.168pt}{0.400pt}}
\put(181,345.67){\rule{2.168pt}{0.400pt}}
\multiput(185.50,346.17)(-4.500,-1.000){2}{\rule{1.084pt}{0.400pt}}
\put(190.0,347.0){\rule[-0.200pt]{1.927pt}{0.400pt}}
\sbox{\plotpoint}{\rule[-0.500pt]{1.000pt}{1.000pt}}%
\put(1320,778){\makebox(0,0)[r]{Imag. part}}
\multiput(1340,778)(20.756,0.000){5}{\usebox{\plotpoint}}
\put(1440,778){\usebox{\plotpoint}}
\put(1054,607){\usebox{\plotpoint}}
\put(1054.00,607.00){\usebox{\plotpoint}}
\put(1073.43,599.83){\usebox{\plotpoint}}
\put(1092.90,592.64){\usebox{\plotpoint}}
\put(1112.48,585.84){\usebox{\plotpoint}}
\put(1132.24,579.50){\usebox{\plotpoint}}
\put(1151.76,572.47){\usebox{\plotpoint}}
\put(1171.62,566.46){\usebox{\plotpoint}}
\put(1191.49,560.50){\usebox{\plotpoint}}
\put(1211.52,555.16){\usebox{\plotpoint}}
\put(1231.42,549.32){\usebox{\plotpoint}}
\put(1251.36,543.69){\usebox{\plotpoint}}
\put(1271.43,538.52){\usebox{\plotpoint}}
\put(1291.50,533.35){\usebox{\plotpoint}}
\put(1311.72,528.73){\usebox{\plotpoint}}
\put(1331.82,523.64){\usebox{\plotpoint}}
\put(1352.13,519.37){\usebox{\plotpoint}}
\put(1372.21,514.18){\usebox{\plotpoint}}
\put(1392.56,510.09){\usebox{\plotpoint}}
\put(1412.87,505.83){\usebox{\plotpoint}}
\put(1433.22,501.73){\usebox{\plotpoint}}
\put(1453.54,497.49){\usebox{\plotpoint}}
\put(1473.86,493.29){\usebox{\plotpoint}}
\put(1480,492){\usebox{\plotpoint}}
\sbox{\plotpoint}{\rule[-0.200pt]{0.400pt}{0.400pt}}%
\put(1053,163){\usebox{\plotpoint}}
\multiput(1053,163)(0.000,20.756){34}{\usebox{\plotpoint}}
\put(1053,859){\usebox{\plotpoint}}
\sbox{\plotpoint}{\rule[-0.500pt]{1.000pt}{1.000pt}}%
\put(1315,365){\raisebox{-.8pt}{\makebox(0,0){$\diamond$}}}
\put(1272,391){\raisebox{-.8pt}{\makebox(0,0){$\diamond$}}}
\put(1228,425){\raisebox{-.8pt}{\makebox(0,0){$\diamond$}}}
\put(1184,469){\raisebox{-.8pt}{\makebox(0,0){$\diamond$}}}
\put(1141,535){\raisebox{-.8pt}{\makebox(0,0){$\diamond$}}}
\put(1097,638){\raisebox{-.8pt}{\makebox(0,0){$\diamond$}}}
\put(1036,790){\raisebox{-.8pt}{\makebox(0,0){$\diamond$}}}
\put(1010,670){\raisebox{-.8pt}{\makebox(0,0){$\diamond$}}}
\put(1075,802){\raisebox{-.8pt}{\makebox(0,0){$\diamond$}}}
\put(966,583){\raisebox{-.8pt}{\makebox(0,0){$\diamond$}}}
\put(923,535){\raisebox{-.8pt}{\makebox(0,0){$\diamond$}}}
\put(879,502){\raisebox{-.8pt}{\makebox(0,0){$\diamond$}}}
\put(835,477){\raisebox{-.8pt}{\makebox(0,0){$\diamond$}}}
\put(792,458){\raisebox{-.8pt}{\makebox(0,0){$\diamond$}}}
\put(748,442){\raisebox{-.8pt}{\makebox(0,0){$\diamond$}}}
\put(704,428){\raisebox{-.8pt}{\makebox(0,0){$\diamond$}}}
\put(661,417){\raisebox{-.8pt}{\makebox(0,0){$\diamond$}}}
\put(617,407){\raisebox{-.8pt}{\makebox(0,0){$\diamond$}}}
\put(574,398){\raisebox{-.8pt}{\makebox(0,0){$\diamond$}}}
\put(530,390){\raisebox{-.8pt}{\makebox(0,0){$\diamond$}}}
\put(486,382){\raisebox{-.8pt}{\makebox(0,0){$\diamond$}}}
\put(443,376){\raisebox{-.8pt}{\makebox(0,0){$\diamond$}}}
\put(399,370){\raisebox{-.8pt}{\makebox(0,0){$\diamond$}}}
\put(355,364){\raisebox{-.8pt}{\makebox(0,0){$\diamond$}}}
\put(312,359){\raisebox{-.8pt}{\makebox(0,0){$\diamond$}}}
\put(268,354){\raisebox{-.8pt}{\makebox(0,0){$\diamond$}}}
\put(225,350){\raisebox{-.8pt}{\makebox(0,0){$\diamond$}}}
\put(181,346){\raisebox{-.8pt}{\makebox(0,0){$\diamond$}}}
\sbox{\plotpoint}{\rule[-0.600pt]{1.200pt}{1.200pt}}%
\put(1054,607){\raisebox{-.8pt}{\makebox(0,0){$\diamond$}}}
\put(1087,595){\raisebox{-.8pt}{\makebox(0,0){$\diamond$}}}
\put(1119,584){\raisebox{-.8pt}{\makebox(0,0){$\diamond$}}}
\put(1151,573){\raisebox{-.8pt}{\makebox(0,0){$\diamond$}}}
\put(1184,564){\raisebox{-.8pt}{\makebox(0,0){$\diamond$}}}
\put(1216,553){\raisebox{-.8pt}{\makebox(0,0){$\diamond$}}}
\put(1248,544){\raisebox{-.8pt}{\makebox(0,0){$\diamond$}}}
\put(1280,536){\raisebox{-.8pt}{\makebox(0,0){$\diamond$}}}
\put(1312,528){\raisebox{-.8pt}{\makebox(0,0){$\diamond$}}}
\put(1345,519){\raisebox{-.8pt}{\makebox(0,0){$\diamond$}}}
\put(1377,514){\raisebox{-.8pt}{\makebox(0,0){$\diamond$}}}
\put(1409,507){\raisebox{-.8pt}{\makebox(0,0){$\diamond$}}}
\put(1442,501){\raisebox{-.8pt}{\makebox(0,0){$\diamond$}}}
\put(1474,494){\raisebox{-.8pt}{\makebox(0,0){$\diamond$}}}
\end{picture}

 \caption{\label{fig:twoptlim}Comparison of our method with a calculation of the
scalar sunset-graph with two squared propagators of mass \(m\). 
(Diamonds represent numerical results from our calculation.)} 
\end{center}
\end{figure}

We compare this with numerical results obtained via another
method~\cite{PostTausk} and find agreement below and above threshold
(figure~\ref{fig:twoptlim}). (For definiteness, all masses have been
chosen equal, requiring yet another expansion due to the
denominators in~(\ref{T2entwickl}).)

\section{Conclusion}

The method for calculating scalar 2-loop box-functions with arbitrary
internal masses proposed in~\cite{hepph9407234} turns out to deliver a
moderate number of 4-dimensional integrals, which can always be
reduced further to 3-dimensional representations---in some cases even
2-dimensional ones. In sample-cases we have been able to produce
reasonable numerical results in arbitrary kinematical regimes below
and above threshold. In the limit of limiting kinematical points,
numerical stability is lost but can be restored by expanding the
representation around that point.  We hope to obtain similar results
for all the 5 genuine 2-loop box-functions and incorporate them into
\verb|xloops|~\cite{xLoopsIntro}.

\section{Acknowledgements}

R.~Kreckel is grateful to the `Graduiertenkolleg
Elementarteilchenphysik bei hohen und mittleren Energien' at
University of Mainz for supporting part of this work. D.~Kreimer
thanks Bob Delbourgo and the Physics Dept. at the Univ. of Tasmania
for hospitality during a visit in March 1998 and the DFG for support.

\end{document}